\documentclass[letterpaper,times]{sae}

\usepackage[svgnames]{xcolor}
\definecolor{SAEblue}{RGB}{1,160,233}
\raggedright

\usepackage[justification=justified, singlelinecheck=false]{caption}  
\DeclareCaptionFont{eightpt}{\fontsize{8pt}{11pt}\selectfont #1}
\usepackage{lineno}
\usepackage[utf8]{inputenc}

\usepackage{array}
\newcolumntype{L}[1]{>{\raggedright\let\newline\\\arraybackslash\hspace{0pt}}p{#1}}
\newcolumntype{C}[1]{>{\centering\let\newline\\\arraybackslash\hspace{0pt}}p{#1}}
\newcolumntype{R}[1]{>{\raggedleft\let\newline\\\arraybackslash\hspace{0pt}}p{#1}}

\usepackage{graphicx}
\usepackage{multirow}
\usepackage{amsmath}

\newcommand{\ignore}[1]{}

\renewcommand{\deg}{$^\circ$}

\usepackage{nameref}
\usepackage{booktabs}

\makeatletter
\def\@seccntformat#1{%
  \expandafter\csname c@#1\endcsname\c@section
  }
\makeatother

\usepackage{subfigure}
\usepackage{multimedia}

\PaperTitle{A Comparison of Road Grade Preview Signals from Lidar and Maps}
\usepackage{calc} 

\usepackage{amssymb}
\usepackage{amsmath}
\usepackage{mathtools}
\usepackage{algorithm}
\usepackage{algpseudocode}
\usepackage{svg}
\usepackage{url}

\makeatletter 
\renewcommand\@biblabel[1]{#1. } 
\makeatother

\AddAuthor{Logan Schexnaydre, Aman Poovalappil, Darrell Robinette, Jeremy Bos}{Michigan Technological University}
\PaperNumber{2019-XX-XXXX}
\SAECopyright{2019}

\begin{document}
\maketitle
\section{Abstract}
Road grade can impact the energy efficiency, safety, and comfort associated with automated vehicle control systems. Currently, control systems that attempt to compensate for road grade are designed with one of two assumptions. Either the grade is only known once the vehicle is driving over the road segment through proprioception, or complete knowledge of the oncoming road grade is known from a pre-made map. Both assumptions limit the performance of a control system, as not having a preview signal prevents proactive grade compensation, whereas relying only on map data potentially subjects the control system to missing or outdated information. These limits can be avoided by measuring the oncoming grade in real-time using on-board lidar sensors. In this work, we use point returns accumulated during travel to estimate the grade at each waypoint along a path. The estimated grade is defined as the difference in height between the front and rear wheelbase at a given waypoint. Kalman filtering techniques are used to mitigate the effects of odometry and motion uncertainty on the grade estimates. This estimator's performance is compared to the measurements of a map created with a GNSS/INS system via a field experiment. When compared to the map-based system, the lidar-based estimator produces an unbiased error with a standard deviation of 0.6 degrees at an average range of 52.7 meters. By having similar precision to map-based systems, automotive lidar-based grade estimation systems are shown to be a valid approach for measuring road grade when a map is unavailable or inaccurate. In using lidar as an input signal for grade-based control system tasks, autonomous vehicles achieve higher redundancy and independence in contrast to existing methods.



\section{Introduction}
Knowledge of the grade of a road can be used to improve the safety, energy savings, and ride comfort of autonomous vehicles and automated vehicle features. For adaptive cruise control systems, accounting for the road grade both increases the precision of safe spacing between vehicles and the ego vehicle's energy efficiency \cite{firoozi_safe_2019}. On-road experiments have been shown to reduce the fuel consumption of a truck by 8\% by reducing gear shifting and enabling proactive velocity profiling over hilly terrain \cite{hellstrom_look-ahead_2009}\cite{grewal_improving_2023}. Simulation results show that hybrid electric vehicles can optimize their energy use strategies based on a road grade signal \cite{zhang_role_2010}. Active suspension systems can proactively adjust their stiffness based on oncoming road grade to reduce uncomfortable cabin motion \cite{ni_road_2020}. All of these systems require some source of road grade knowledge. 

Grade information can be created either as the vehicle experiences it or before the vehicle reaches that road segment. The first type of system, when used with proprioceptive sensors, lets a vehicle perform reactive tasks, such as calculating the mass and road load of a vehicle \cite{bae_road_2001} \cite{poovalappil_real-time_2025}. With these parameters known, a vehicle can be safely controlled to the correct speed, or a vehicle's current maximum range can be calculated. These systems often estimate the road grade using a combination of Global Navigation Satellite System (GNSS) and Inertial Navigation System (INS) information in a Kalman filter \cite{gokcek_novel_2020}\cite{jauch_road_2018}. However, knowing only the current road grade does not let a vehicle prepare for oncoming grades, which limits improvements in vehicle performance.

A preview of the oncoming road grade can enable a control system to manage hilly terrain proactively. Complete knowledge of the oncoming path grade is often assumed to be known for control systems that require grade information \cite{firoozi_safe_2019} \cite{hellstrom_look-ahead_2009}. Maps of the road grade preview can be made via GNSS and vehicle powertrain sensors by driving a vehicle over the road segment before the test run \cite{sahlholm_road_2010}. A preview of road grade can also be provided by connected leading vehicles as well \cite{ma_cooperative_2020}. Digital elevation maps (DEMs) can be generated from airborne \cite{zhang_road_2006} and terrestrial lidar \cite{gargoum_fully_2018}. However, maps like these face challenges in being produced for large regions, remaining current, and being communicated to an autonomous vehicle (AV) \cite{wong_mapping_2021}. When this map information is unobtainable or imprecise, proactive grade-based control systems will perform suboptimally.

Automotive lidar sensors have the potential to provide grade measurements in place of a map. These sensors provide spatial information about the environment that can be processed to measure the elevation profile of a road \cite{ni_road_2020}. The derivative of the elevation profile can be used to create a grade profile. When subject to odometry noise during travel, the uncertainty in this derivative estimation can become significant. Kalman filtering has been performed on the GNSS/INS measurements of a lidar-based elevation profiling system \cite{zhao_extraction_2018} and directly on the elevation map \cite{wang_extraction_2020}, but the performance of a lidar-based grade estimator system relative to a grade map has yet to be understood.

We present the results of field experiments conducted to compare the accuracy and precision of automotive lidar-based grade measurements relative to an inertial map-based system. The lidar-based approach estimates a vehicle's experienced grade using the height difference between point returns located at the front and rear wheel contact patches. An odometry bias dependent on the order in which the front and rear patches acquire measurements is corrected. A Kalman filter is used to mitigate the effect of GNSS/INS uncertainty on the lidar-based grade measurements. Field experiments are conducted to compare the lidar-based estimator to an INS-based map of road grade along the traveled path. The experiment results indicate that the lidar-based estimator is unbiased and has similar precision relative to the map-based system. The computational complexity of the lidar processing algorithm is identified as linear and directly proportional to the maximum range and resolution of the grade estimates being generated.

\section{Related Works}
Maps of grade are created using methods described previously, such as recording prior measurement of the change in elevation of a vehicle's GNSS \cite{sahlholm_road_2010}, or the pitch of inertial sensors \cite{jauch_road_2018}. These approaches measure the road grade at a set of waypoints along the path traveled. DEMs generated from aerial \cite{zhang_road_2006} or survey-grade terrestrial \cite{gargoum_fully_2018} lidar are used to extract grade measurements in a similar path format, but they require post-processing of the DEM. An automotive lidar-based system must create measurements in the same manner to enable comparison of its precision and accuracy with map-based systems.

Ground plane height estimation can be a prerequisite to grade estimation and is also a component of the ground and road segmentation tasks. GNDNet is based on PointPillars \cite{lang_pointpillars_2019} and estimates the DEM of the ground by binning its points along the X-Y coordinate plane, creating a pseudoimage from those bins, and estimating the elevation map of the ground using a neural network \cite{paigwar_gndnet_2020}. The grade of a road could then be calculated from the change in height of the DEM, provided road segmentation occurs. PLARD is another neural network, but it is used for road segmentation \cite{chen_progressive_2019}. This approach converts the point returns to an image space using each point's height coordinate, then applies a neural network to fuse the road features extracted from the camera and the lidar. However, PLARD only outputs labels of road segments for an image and does not retain elevation information. To provide a representative example of a lidar-based grade estimator for this comparison, our approach bins points similarly to PointPillars, but only within the front and rear wheel contact patch regions for each waypoint.

Grade estimation is often performed in lidar-based ground segmentation for mobile robots. These systems typically segment the ground using an estimate of the normal vector of the surface defined by a set of neighboring points. Patchwork \cite{lim_patchwork_2021} extracts the normal vector of the plane fit to a set of points through Principal Component Analysis (PCA). This normal vector is then used in combination with elevation and roughness criteria to label the points as ground or otherwise. A least squares approach is another approach to estimating the normal of a surface and has been applied to grade estimation \cite{alzubi_lidar-based_2020}. PCA and least squares have been shown to have similar performances in normal estimation \cite{jordan_quantitative_2014}. However, this surface fitting method may not reflect the grade experienced by the vehicle at a given waypoint, as the point set could include points not within the vehicle's contact regions. For a comparison of a map-based and lidar-based approaches, we make a similar planar assumption, but only use the elevation of points in contact patch regions to create the grade.

\section{Method}
\begin{figure}[!htb]
\centering
\includegraphics[width=0.35\textwidth]{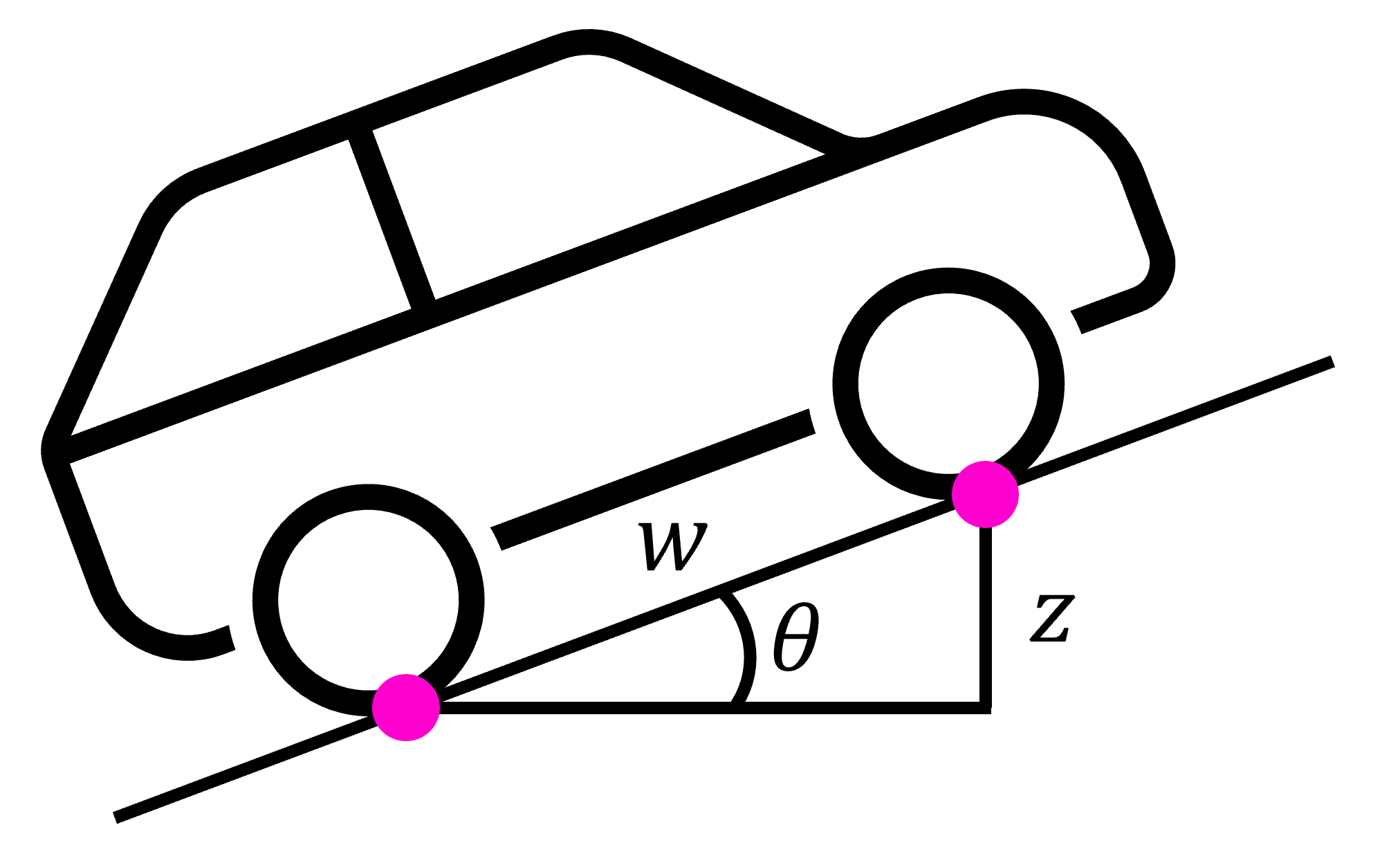}
\caption{At each waypoint, the difference in height between the front and rear contact patch, $z$, and the wheelbase length $w$, can be used to estimate the experienced grade $\theta$.}
\label{fig:grade_diagram}
\end{figure}

The instantaneous grade that a vehicle experiences at a given waypoint is defined using the relative difference in height between the front wheels and rear wheels, $z \in \mathbb{R}$, and the constant wheelbase length, $w\in \mathbb{R}^+$. These values form a right triangle as shown in Figure \ref{fig:grade_diagram}. The grade that a vehicle experiences at that waypoint is then
\begin{equation}  \label{eq:grade}
    \theta = \sin^{-1}\frac{z}{w}.
\end{equation} 
The grade can be estimated for a set of waypoints that define the vehicle's pose along the path $S^W = \{s_i:i=1,\dots,M\}, s_i=\{x_i,y_i,z_i,\phi_i,\theta_i,\psi_i\}$, in the world frame $W$.  The waypoints are uniformly spaced apart at a distance $\Delta s$.

A lidar-based grade estimation system must be developed to create equivalent estimates of the road grade. Due to the limited range of a lidar sensor, the system will be set to measure grade to some maximum range ahead of the vehicle, $d \in \mathbb{R}^{>0}$. This maximum range indicates for which points in $S^W$ a grade measurement can be made.

\begin{figure}[h]
    \centering
    \includegraphics[width=1\linewidth]{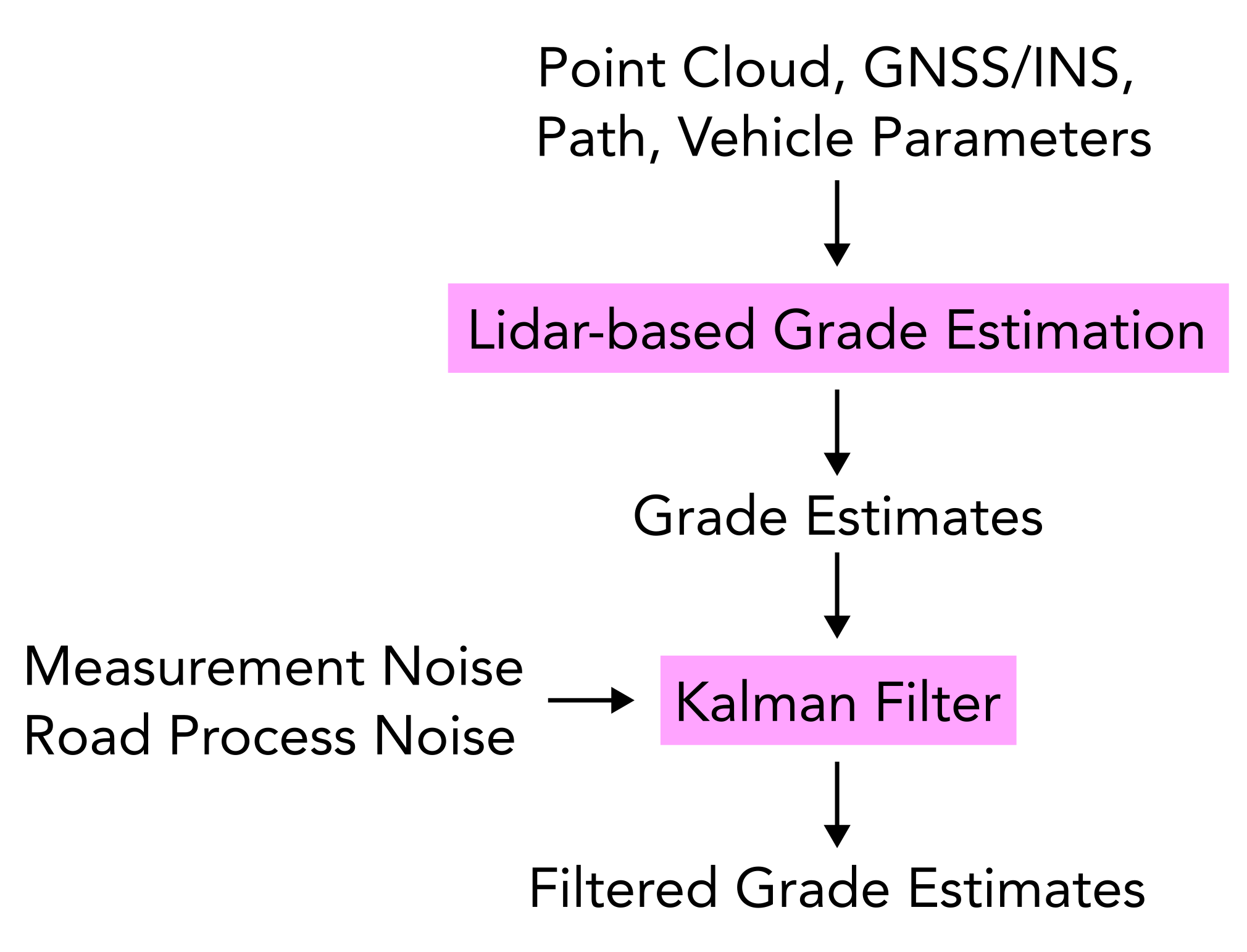}
    \caption{The process used to create filtered grade estimates. The magenta boxes are the two major steps in the grade estimation process: grade estimation and Kalman filtering.}
    \label{fig:flowchart}
\end{figure}

The two main steps of the lidar-based grade estimation system are the estimation and Kalman filtering steps. Figure \ref{fig:flowchart} depicts the process. Lidar point returns, the transform of the lidar frame to the world frame (from GNSS/INS measurements), the path, and the vehicle's shape are all inputs to the grade estimation system. Within this system, the lidar point returns are used to determine the wheel heights at each waypoint and create the road grade estimate. For each set of point returns that this system processes, points at each contact patch are accumulated and subject to odometry noise. These estimates with noise are passed to the Kalman filter to reduce the effects of odometry uncertainty. The road process noise and the grade measurement noise are supplied to the Kalman filter to complete the estimate of the grade state.

\subsection{Lidar-based Grade Estimation} 
A lidar's data product is defined as a set of point returns, $P_t^L = \{p_i : i=1,\dots,N\},~p_i=\{x_i,y_i,z_i\}$, at time $t$ in the lidar coordinate frame, $L$. The front and rear wheel contact patches at $s_i$ are defined as 3D rectangular regions with position, yaw, and constant size
\begin{equation}
    c_{f,i} = \{x_i+\Delta x/2,y_i+\Delta y/2,z_i,\psi_i,l,w_c,\infty\}
\end{equation} and 
\begin{equation}
    c_{r,i} = \{x_i-\Delta x/2,y_i-\Delta y/2,z_i,\psi_i,l,w_c,\infty\}.
\end{equation} $\Delta x$ and $\Delta y$ are the distance components from the front contact patch to the rear contact patch of the rotated vehicle, $\Delta x=w\,\sin(\psi_i)$ and $\Delta y = w\,\cos(\psi_i)$. The distance between the left and right wheels for either contact patch is $l\in\mathbb{R}^+$. These patches are included in the contact patch set $C^W$. The roll and pitch of these regions are assumed to be zero, as they are used as box filters to select only point returns that will inform the height of the wheels. The height of the contact patch box is set to infinity to avoid an incorrect assumption on the road elevation at a given waypoint. Another set of box filters can be applied after this first set to filter around the minimum points gathered from the first filter.

\begin{figure}[h]
    \centering
    \includegraphics[width=1\linewidth]{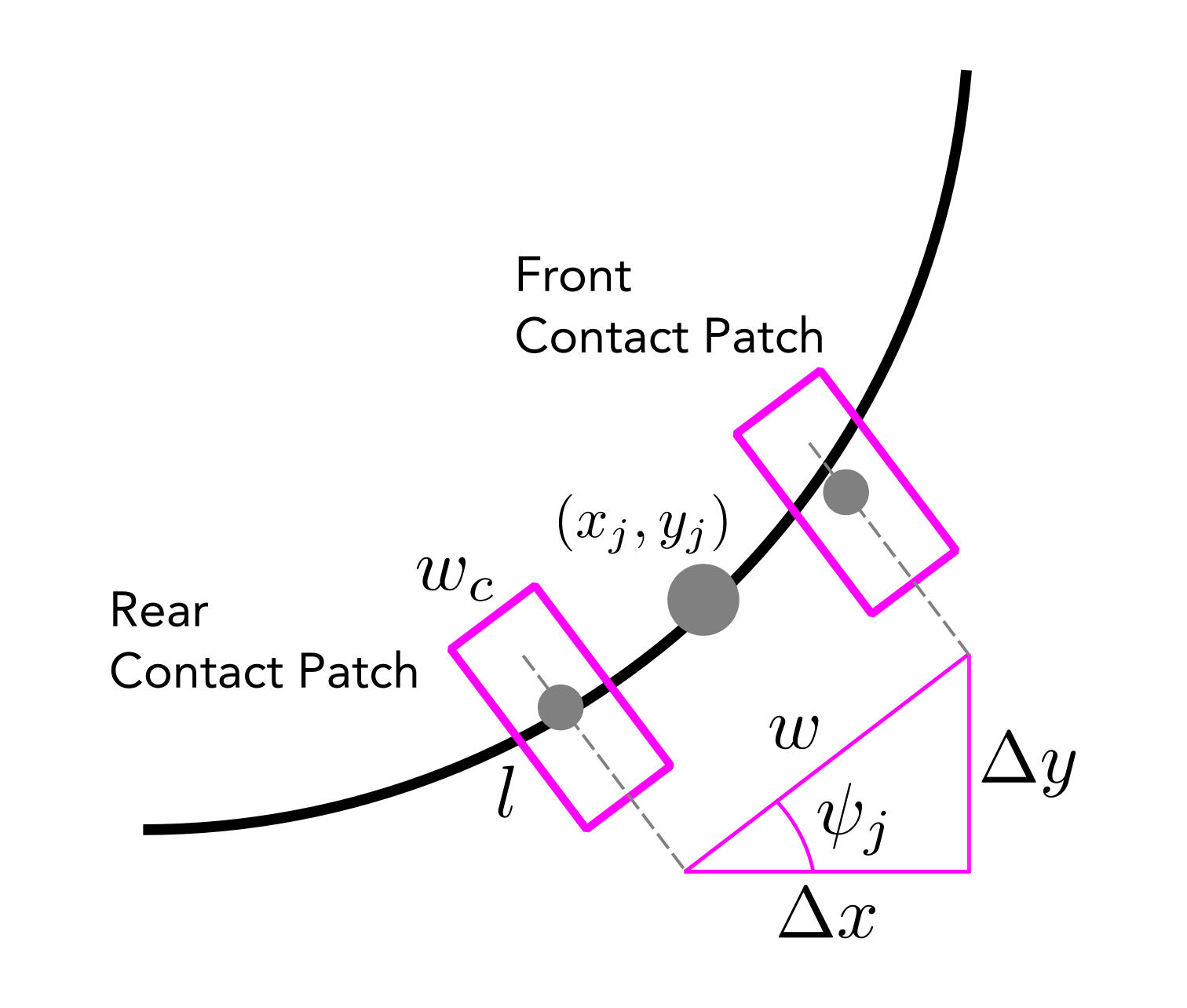}
    \caption{A bird's-eye view of the vehicle front and rear contact patch locations when a vehicle is centered at a waypoint with a position $(x_j,y_j)$. The contact patches are equidistant from the vehicle centerpoint and rotated about that point with an angle $\psi_j$.}
    \label{fig:contact_patches}
\end{figure}

Each set of point returns will first be transformed from the lidar frame to the world frame using the time-varying transform, $T^L_W(t)$. With $P_t^W$ and the current vehicle position known, all front and rear contact patches for each waypoint ahead of the vehicle will be used as box filters to determine if a point return is on a contact patch. If so, then the point is recorded as a height measurement for that contact patch. Once a contact patch has at least one point registered, it no longer accumulates points. The other patch will continue to search for points until at least one point can be stored in the point cloud structures $Q^W_{f,i}=\{q_{f,i} : i=1,\dots,W\},~q_{f,i}=\{x_{f,i},y_{f,i},z_{f,i}\}$ or $Q^W_{r,i}=\{q_{r,i} : i=1,\dots,V\},~q_{r,i}=\{x_{r,i},y_{r,i},z_{r,i}\}$. Once at least one measurement is present in both front and rear contact patches, the difference in height between the front and rear wheels is estimated as the difference in average height of the points in front and rear contact patches
\begin{equation} \label{eq:height_estimate}
\hat{z}_i = \frac{1}{W}\sum_{k=1}^{W}z_{f,k} - \frac{1}{V}\sum_{k=1}^{V}z_{r,k}.
\end{equation}
This is used to produce the grade estimate for waypoint $s_i$
\begin{equation}
\hat{\theta}_i = \sin^{-1}\frac{\hat{z}_i}{w} + b(\Delta f_i).
\end{equation}
The function $b(\Delta f_i)$ is to correct for bias that can occur due to accumulated odometry error \cite{martinelli_estimating_2003}, where $\Delta f_i \in \mathbb{Z}$ is the number of frames elapsed between when the first measurements occur within the front and rear contact patches. A positive $\Delta f_i$ indicates that the front contact patch received measurements first; a negative $\Delta f_i$ indicates the opposite. The bias correction will be defined as the piecewise function of linear models
\begin{equation}
    b(\Delta f_i)=\begin{cases}
m_f\Delta f_i + b_f & \text{if } \Delta f_i > 0 \\
m_r|\Delta f_i| + b_r  & \text{if } \Delta f_i < 0 \\
0& \text{Otherwise}
\end{cases}.
\end{equation}

These measurements are subject to noise from the range uncertainty of the lidar and the uncertainty in $T_W^L(t)$ from the vehicle odometry system during travel. While the range uncertainty is independent across all point returns, the measurement bias and variance depend on the amount of time odometry uncertainty is allowed to accumulate between the front and rear contact patches receiving their first measurements. 

This process is summarized in Algorithm \ref{alg:cap}. The set of grade estimates is initialized as an $M$-length set of null entries to distinguish between waypoints with and without an estimate $\hat\theta = \mathbf{0}_M$.

\begin{algorithm}
\caption{Grade Estimation}\label{alg:cap}
\begin{algorithmic}
\Require \\$P_t^L$, $S^W$, $T_W^L$, $C_W$, $Q^W$\\
         $i$ \Comment{Closest waypoint index}\\
         $d$ \Comment{Maximum preview distance}\\
         $\Delta s$ \Comment{Distance sample rate}\\
\Ensure $\hat\theta^W$, $Q^W$\\

\State $P_t^W \gets \text{TransformPointCloud}(P_t^L,T_W^L)$
\For{$j = i~\text{to}~\left\lfloor\frac{d}{\Delta s}\right\rfloor$}
    \State $P_{t,f}^W \gets \text{BoxFilter}\left(P_t^W,C_{f,j}^W\right)$
    \State $P_{t,r}^W \gets \text{BoxFilter}\left(P_t^W,C_{r,j}^W\right)$ 
    \If{ $|Q^W_{f,j}| = 0$} \Comment{Accumulate points in each contact patch}
        \State $Q^W_{f,j} = Q^W_{f,j} + P^W_{t,f}$
        \State $f_f = \text{GetFrameNumber}()$
    \EndIf
    \If{ $|Q^W_{r,j}| = 0$}
        \State $Q^W_{r,j} = Q^W_{r,j} + P^W_{t,r}$
        \State $f_r = \text{GetFrameNumber}()$
    \EndIf
    
    \If{ $|Q^W_{f,j}| \geq 1 \wedge |Q^W_{r,j}| \geq 1$}
        \State $\Delta f_i = f_f - f_r$
        \State $\mu_f \gets \frac{1}{N}\sum_{k=1}^{N}z_{f,k}$
        \State $\mu_r \gets \frac{1}{V}\sum_{k=1}^{V}z_{r,k}$
        \State $\hat\theta^W_i \gets \sin^{-1} \frac{\mu_f - \mu_r}{w} + b(\Delta f_i)$
    \EndIf
\EndFor

\end{algorithmic}
\end{algorithm}

It is useful to know the computational complexity of this Algorithm \ref{alg:cap} to adjust its parameters for real-time use on a vehicle. The coordinate transformation of the point cloud and box filter steps both must iterate through all points in $P_t$ once, resulting in each step being within the complexity $O(N)$. As most of $P_{t}^W$ is not likely to be within either box filter, we can assume $|P_{t,f}^W| \ll N$ and $|P_{t,r}^W| \ll N$. It is also not likely that $Q_{f,j}^W$ or $Q_{r,j}^W$ will be a comparable size to $P_t^W$. Because of these reasons, the two-box filter steps drive the cost of computing the grade for one waypoint (the inside of the for loop) to be $O(N)$. 

When considering the number of waypoints within the maximum preview distance, $\left\lfloor\frac{d}{\Delta s}\right\rfloor$, and the two box filters, the computational complexity of the for loop becomes $O\left(\frac{2Nd}{\Delta s}\right)$. Combining this with the coordinate transformation results in $O\left(N+\frac{2Nd}{\Delta s}\right)$. Removing constants and considering which term drives the runtime cost of Algorithm \ref{alg:cap} to be in the computational complexity 
\begin{equation}
g(N,d,\Delta s) \in O\left(N\cdot\text{max}\left(1,\frac{d}{\Delta s}\right)\right).
\end{equation}
This computational complexity directly relates the demand of the lidar processing component of the grade estimation system to the maximum estimation range and the number of waypoints within that range.



\subsection{Kalman Filtering}

A Kalman filter will be used on the lidar-based grade measurements in a spatially-indexed domain, $s_i$, to mitigate the effects of odometry uncertainty. 
A constant grade velocity model is used to estimate $\theta$ using the measurements $\hat\theta_i$. The states to be estimated are
\begin{equation}
    \mathbf{x}[s] =
    \begin{bmatrix}
    \theta \\
    \dot \theta
    \end{bmatrix}.
\end{equation}
There are no control inputs modeled to affect $\mathbf{x}$. The constant grade velocity model state transition matrix is set as
\begin{equation}
    \Phi=
    \begin{bmatrix}
        1 & \Delta s \\
        0 & 1
    \end{bmatrix}.
\end{equation}
Process noise is only modeled on $\dot\theta$, so the noise input matrix is
\begin{equation}
    \mathbf{G} = 
    \begin{bmatrix}
    0 \\ 1
    \end{bmatrix}.
\end{equation}
The process noise is modeled using a scalar value $q\in \mathbb{R}^{\ge 0}$, which indicates the variance of the rate of change of grade. With these variables expressed, the integral term of the state covariance prediction equation can be simplified to 
\begin{equation}
    \int \mathbf{\Phi}\mathbf{G}q\mathbf{G}^T\mathbf{\Phi}^T\, d\Delta s = q
    \begin{bmatrix}
        \frac{\Delta s^3}{3} & \frac{\Delta s^2}{2} \\
        \frac{\Delta s^2}{2} & \Delta s
    \end{bmatrix}.
\end{equation}

The only measurement input to the filter is the lidar-based grade estimator's output $\textbf{z}[s] = [\hat{\theta}_i]$.
Since there is only one measurement, the observation matrix becomes
\begin{equation}
    H = 
    \begin{bmatrix}
    1 & 0
    \end{bmatrix}
\end{equation}
and the measurement matrix can be simplified to a scalar value
\begin{equation}
    R(s_i)= \sigma_r^2(s_i) \in \mathbb{R}^{\ge 0}.
\end{equation}
This measurement variance is the summation of several independent measurement Gaussian noise sources. Odometry noise that affects the translational and rotational components of $T_W^L$ can be modeled as such \cite{martinelli_estimating_2003}. These components can be accounted for within $\sigma_r^2$.

The effect of odometry bias is corrected by $b(\Delta f_i)$, but the variance can be addressed within the Kalman filter. We assume the increase in variance of the odometry system with increasing $|\Delta f_i|$ is small enough to create a constant model of variance, resulting in the summation of a constant lidar range uncertainty and odometry uncertainty.
\begin{equation}
    \sigma_r^2(s_i) = \sigma_l^2 + \sigma_o^2.
\end{equation}
This Kalman filter is applied to the estimates available for every waypoint in the path. To minimize the number of iterations of filtering steps, the filter is only used from the start of the path to the farthest waypoint that has a grade estimate. 






\section{Experiment}
The lidar-based grade estimator will be compared to a map-based system in a physical experiment. A Chrysler Pacifica was equipped with a Velodyne VLP-32 lidar and a NovAtel PwrPak7 GNSS/INS receiver. The transform $T^L_W$ is modeled as the combination of the static transformation from the lidar to the GNSS frame and the time-varying transformation of the GNSS frame to the world frame, $T^L_W(t_i)=T^L_GT^G_W(t_i)$.

\urldef\novatelpath\path{https://docs.novatel.com/OEM7/Content/Technical_Specs_Receiver/PwrPak7_IMU_Specifications.htm?tocpath=Specifications%7CPwrPak7%20technical%20specifications%7C_____3}

\begin{figure}[h]
    \centering
    \includegraphics[width=1\linewidth]{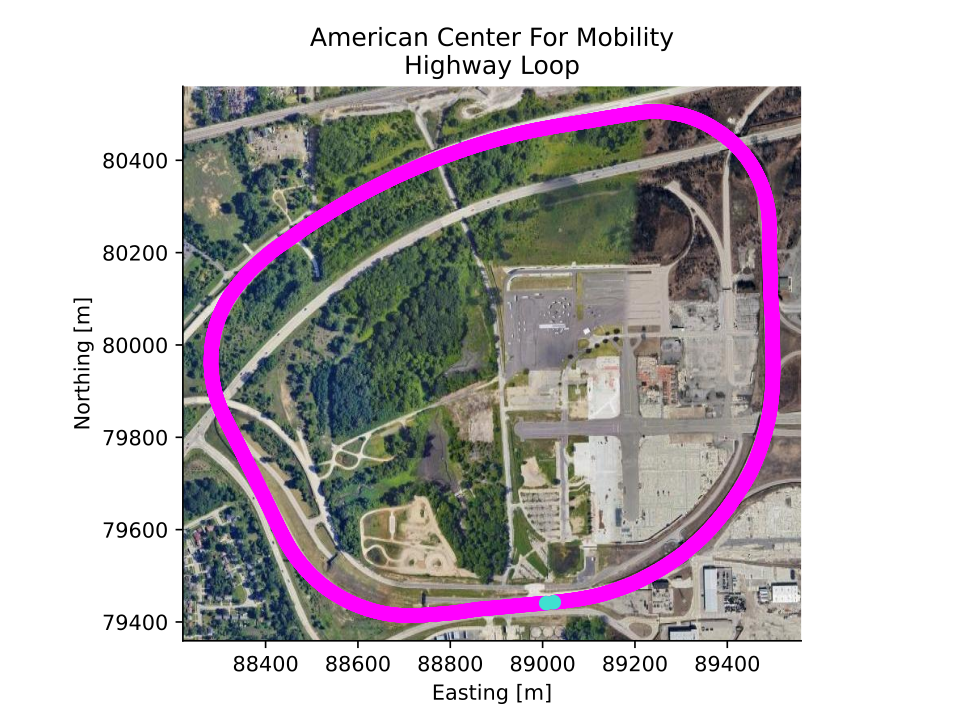}
    \caption{The path recorded by the NovAtel PwrPak7 is a single loop where the blue dots are the start and end points of the path. The route contains an overpass, an underpass, and hilly terrain.}
    \label{fig:acm_map}
\end{figure}

\begin{figure}[h]
    \centering
    \includegraphics[width=1\linewidth]{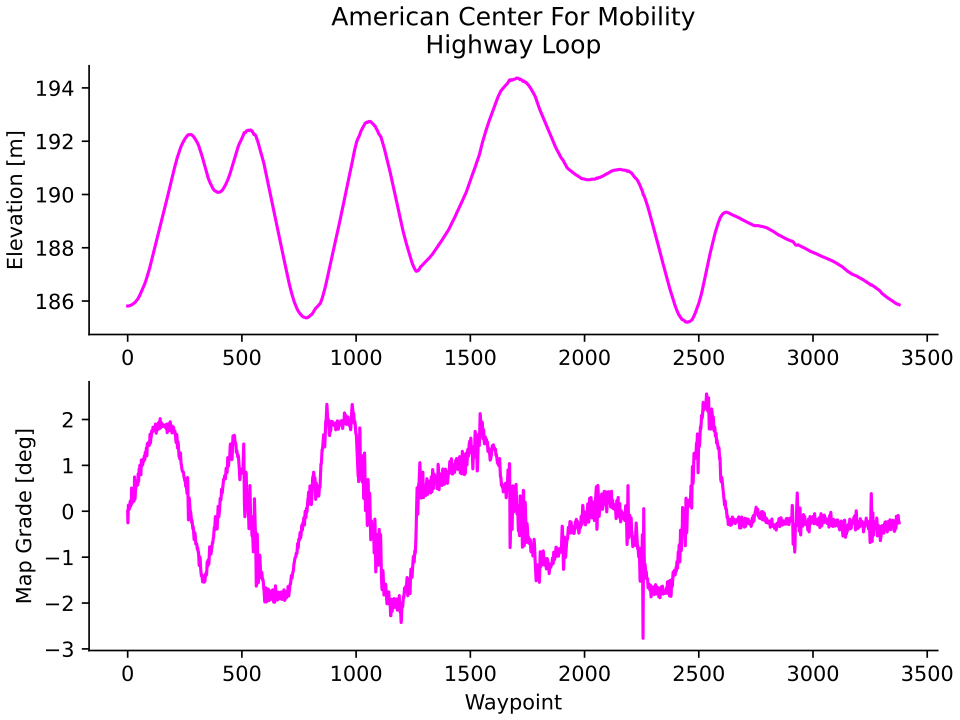}
    \caption{The elevation and grade angle recorded by the NovAtel PwrPak7 at each waypoint. Grades within the range of $\pm2^\circ$ are measured for this highway segment. This grade recording is used as the map-based grade angle for this experiment.}
    \label{fig:acm_grade}
\end{figure}

The experiment was conducted on the American Center for Mobility Highway Loop. The lidar and GNSS/INS measurements were recorded when driving over the track. A reference map-based grade measurement was made by recording the pitch output by the GNSS/INS unit at the interval $\Delta s = 1\text{m}$. The path, elevation, and grade measured by the GNSS/INS unit are shown in Figures \ref{fig:acm_map} and \ref{fig:acm_grade}. A 3.5$^\circ/\text{hr}$ bias instability and a 0.1$^\circ/\sqrt{\text{hr}}$ angular random walk are reported for the IMU gyroscope\footnote{\novatelpath}. The path was traveled in five minutes with a target velocity of 15 meters per second. These points form the path $S$ on which the lidar-based grade estimator predicts the road grade.

The GNSS/INS unit is mounted in the cabin, which is on top of the vehicle's suspension. This suspension will filter out many high-frequency components of the road's effect on the grade \cite{gokcek_novel_2020}\cite{jauch_road_2018}. Because of this, the lidar-based system's grade estimates that are produced from road point returns will have higher frequency components not present in the INS due to the low-pass filtering behavior of the vehicle suspension. The difference between these two signals is accounted for in the Kalman filter's measurement noise and process noise values.

\begin{figure}[h]
    \centering
    \includegraphics[width=1\linewidth]{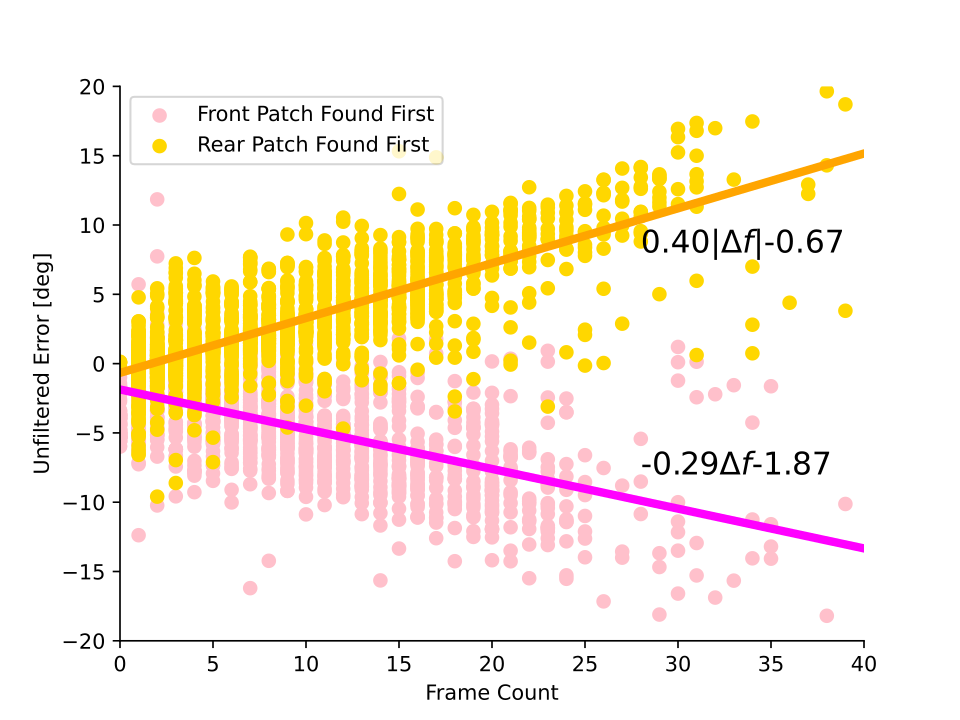}
    \caption{The error of the unfiltered lidar-based grade estimator as the number of frames between front and rear contact patch measurements increases. Separate linear models were fit to the error point in which the first measurement was in the front contact patch or the rear contact patch. The parameters of each fit are used to compensate for bias in the measurements before Kalman filtering occurs.}
    \label{fig:patch_error}
\end{figure}

The parameters set for the estimation process are defined in Table \ref{tab:GE_and_KF_parameters}. The Chrysler Pacifica vehicle dimensions are set in $w$ and $l$. Since there are regions in the VLP-32 scan pattern that are sparse, the contact patch size is set to 0.5m to reduce the number of frames required to have point returns for both the front and rear wheel locations at a waypoint. The road process noise was selected by calculating the variance of the second derivative of the grades recorded in Figure \ref{fig:acm_grade}. The terms $m_f,~m_r,~b_f$, and $b_r$ are measured empirically by inputting the experiment data into the unfiltered lidar-based grade estimator. Figure \ref{fig:patch_error} shows that the unfiltered error was fit to a linear function to identify the bias correction terms. The measurement uncertainty $\sigma_r^2$ was selected experimentally once the bias was corrected. For this experiment, the algorithm was executed on an AMD Ryzen Threadripper 3960x with 64GB of DDR4 memory.

\begin{table}[!htb]
\fontsize{8}{10}\selectfont
\centering
\caption{Grade Estimator and Kalman Filter Parameters.}\label{tab:GE_and_KF_parameters}
\begin{tabular}{| L{0.2\columnwidth-2\tabcolsep-1.2\arrayrulewidth} | L{0.4\columnwidth-2\tabcolsep-1.2\arrayrulewidth} |}
\hline
\textbf{Parameter} & \textbf{Value} \\ \hline
$d$    & 75m \\  \hline
$w$ & 3.09m \\ \hline
$w_c$ & 0.5m \\  \hline
$l$ & 1.73m \\ \hline
$q$ & 8.2e-5 $\text{degrees}^2 /\text{m}^3$ \\ \hline
$\sigma_r^2$ & 49 $\text{degrees}^2$ \\ \hline
$m_f$ & -0.29\deg \\ \hline
$m_r$ & 0.40\deg \\ \hline
$b_f$ & -1.87\deg \\ \hline
$b_r$ & -0.67\deg  \\ \hline
$\Delta s$ & 1m \\ \hline

	\end{tabular}
	\par
   \vspace{-0.15\skip\footins}
   \renewcommand{\footnoterule}{}
\end{table}

\subsection{Results} \label{subsec:results}

\begin{figure}[h]
    \centering
    \includegraphics[width=1\linewidth]{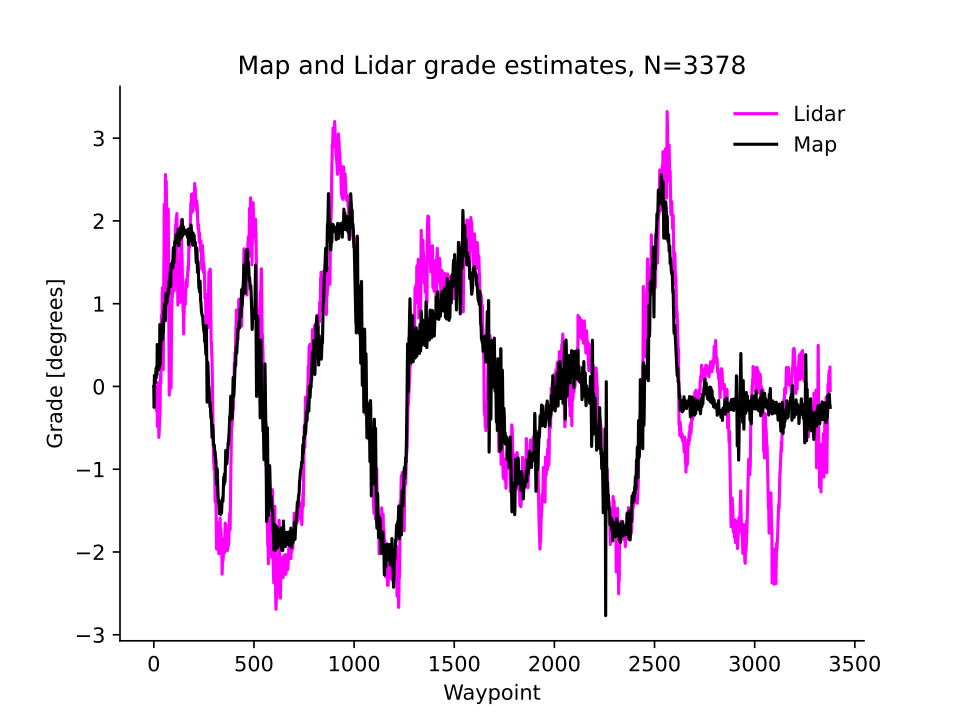}
    \caption{A comparison of the grade estimated by the Kalman-filtered lidar-based estimator and the map generated by INS measurements.}
    \label{fig:estimate_v_truth}
\end{figure}

From this experiment, the precision of the lidar-based system can now be compared to the map generated from INS measurements. Figure \ref{fig:estimate_v_truth} shows a comparison between the two measurement systems. For most waypoints, the lidar-based estimates are similar to the map-based estimates, but have overshoot and undershoot during transient or instantaneous changes in the rate of change of grade. The most significant error occurs towards the end of the route, where the map-based measurements are relatively constant, but the lidar-based measurements vary.

\begin{figure}[h]
    \centering
    \includegraphics[width=1\linewidth]{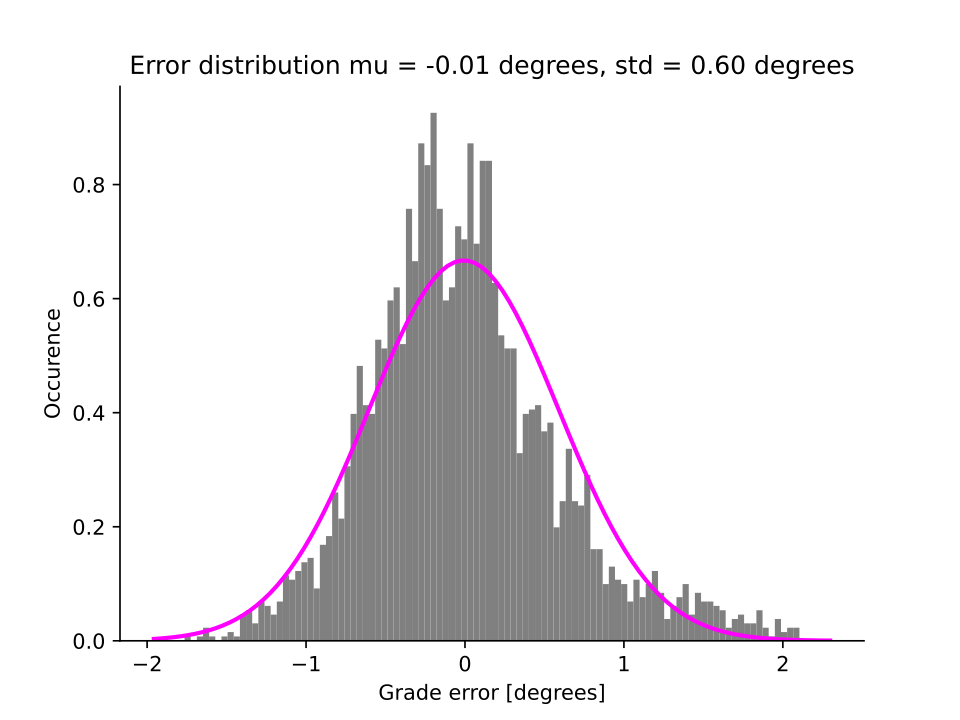}
    \caption{The error in the Kalman-filtered lidar-based estimator when being compared to the map-based measurements. The error fits to a Gaussian distribution with a mean error of -0.01\deg and a standard deviation of 0.60\deg.}
    \label{fig:error_histogram}
\end{figure}

The relative error between the two systems, calculated as the difference between the map-based measurement and the lidar-based measurement, can be seen in Figure \ref{fig:error_histogram}. The error results can be fit to a Gaussian distribution with a mean error of -0.01\deg and a standard deviation of 0.60\deg.

\begin{figure}[h]
    \centering
    \includegraphics[width=1\linewidth]{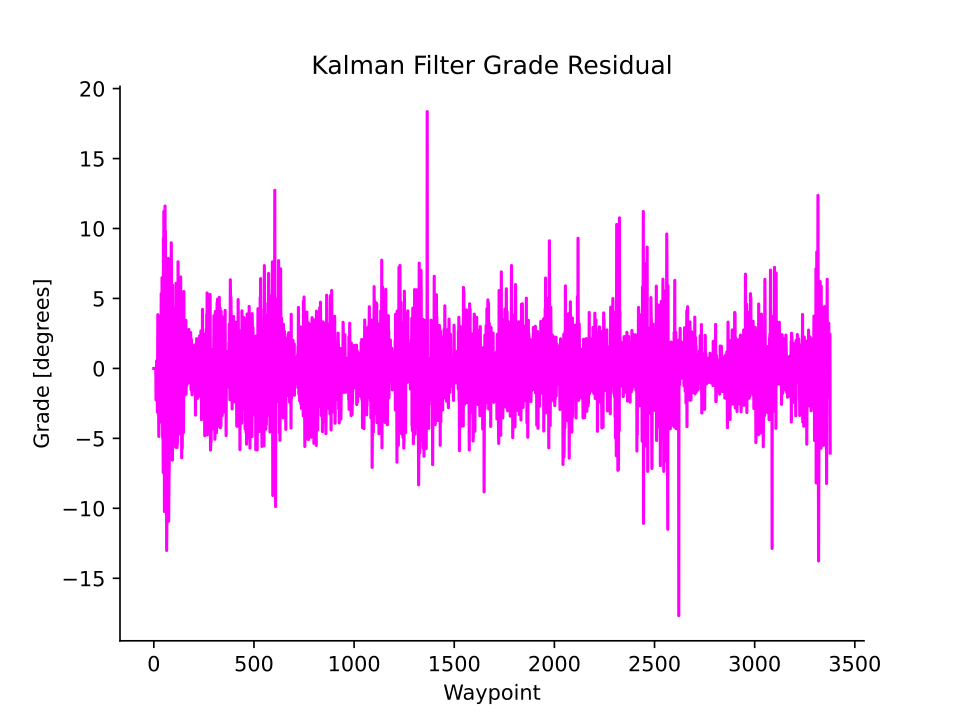}
    \caption{The residual of the Kalman filter applied to the lidar-based grade measurements. The zero mean, small correlation indicates the Kalman filter is closely estimating the grade model dynamics.}
    \label{fig:residual}
\end{figure}

The effect of the Kalman filter can be observed in the filter's residual signal \cite{maybeck_stochastic_1979}. This residual is defined as 
\begin{equation}
\mathbf{r}(s_i) = \mathbf{z}_{i}-\mathbf{H}(s_{i})\hat{\mathbf{x}}(s_{i}^-),
\end{equation}
and, for this experiment, is shown in Figure \ref{fig:residual}. The residual is zero-mean with small autocorrelation, indicating it is close to white noise. This implies that the Kalman filter closely models the dynamics of the map-based grade via the lidar-based sensor.

\begin{figure}[h]
    \centering
    \includegraphics[width=1\linewidth]{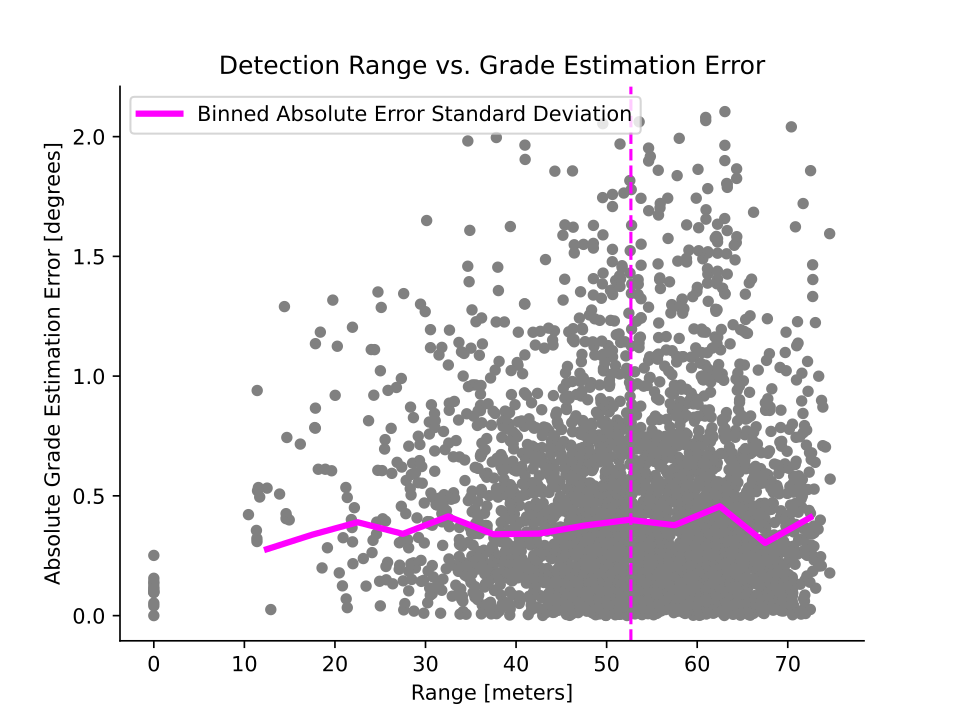}
    \caption{The absolute Kalman-filtered grade estimation error with respect to the range at which the estimate was made. The solid magenta line is the absolute error standard deviation for points binned at 5m intervals. The dotted magenta line is the average range estimate, 52.7m.}
    \label{fig:range_error}
\end{figure}

The range at which the lidar-based grade estimator generates an estimate can be seen in Figure \ref{fig:range_error}. The average range at which an estimate is made is 52.7m, but it approaches the lookahead distance set in the experiment. The binned standard deviation of error in the lidar-based grade estimates relative to the map-based grade estimates is shown by the magenta line. The standard deviation of error remains below 0.5\deg at ranges where there are enough points in the bin to estimate the standard deviation.

Table \ref{tab:times} shows the minimum, average, and maximum execution times of the algorithm. The difference between the minimum and maximum execution times is due to the increasing computational demand of the Kalman filter as the number of waypoints to process with each iteration increases.

\begin{table}[!htb]
\fontsize{8}{10}\selectfont
\centering
\caption{Execution time statistics of the lidar-based grade estimator.}\label{tab:times}
\begin{tabular}{| L{0.22\columnwidth-2\tabcolsep-1.2\arrayrulewidth} | L{0.22\columnwidth-2\tabcolsep-1.2\arrayrulewidth} |L{0.22\columnwidth-2\tabcolsep-1.2\arrayrulewidth} |}
\hline
\textbf{Minimum Execution Time [ms]} & \textbf{Average Execution Time [ms]} & \textbf{Maximum Execution Time [ms]} \\ \hline
7.6 & 58.6 & 112.9 \\  \hline
	\end{tabular}
	\par
   \vspace{-0.15\skip\footins}
   \renewcommand{\footnoterule}{}
\end{table}

\section{Discussion}
This work demonstrates that lidar-based grade estimation systems can estimate road grade with similar precision to a map-based system. The unbiased error being within a standard deviation of 0.60\deg~ (or 1.05\%) between the lidar-based and map-based systems shown in Figure \ref{fig:error_histogram} indicates that lidar-based methods can support grade preview tasks akin to a map-based system. This support could be in place of the map-based grade measurements when they are unavailable or unreliable. There is also potential to fuse each system's measurements to provide information for coarse and fine control input planning. 

The bias compensation shown in shown in Figure \ref{fig:patch_error} and the Kalman filter minimize the effects of odometry uncertainty. This is seen by the minimal change in the absolute error standard deviation with respect to the range an estimate was made in Figure \ref{fig:range_error}. The uncertainty introduced from odometry error will change depending on the quality of the system generating $T^W_L(t)$, and can be accounted for by fitting the bias and adjusting measurement and process covariances in the Kalman filter. The deviation of the lidar-based measurements from the map-based measurements for waypoints 2800-3300 in Figure \ref{fig:estimate_v_truth} likely comes from an unmodeled uncertainty in  $T^W_L(t)$.

If the preview-based control system is using map information to form its original control output, the lidar-based system can be used to update the local control planning. From Figure \ref{fig:range_error}, the lidar-based system's average measurement range is 52.7 meters when traveling at a speed of approximately 15 meters per second. A preview-based control system would have 3.50 seconds to prepare for the oncoming waypoint. This combination of systems would allow for a control system to have a coarse plan for the global mission with fine adjustments once the local region is reached. If the map-based system is unavailable, the controller will also still have a signal ready for local planning.

A limitation of the lidar-based grade estimator is its dependency on the sensor's resolution when sampling the road. If the scan pattern is sparse, then the measurement range decreases as more frames are needed to complete the estimation process. This limitation can be mitigated by reducing the expected value of $\Delta f_i$.  One way $\Delta f_i$ can be reduced is by increasing the size of the contact patches. By doing so, point returns are more likely to be found in less time. The size of the artificial increase needs to be minimal to avoid the risk of using points that do not reflect the height of the tire at a contact patch while minimizing the amount of odometry uncertainty being introduced to a grade estimate. For a higher resolution lidar, the contact patch region sizes would not need to be increased as significantly, as more point returns are likely to be within the true contact patch region.

The computational time of the algorithm can be considerably shortened by reducing the number of waypoints the Kalman filter must process with each step. The minimum time of 7.6ms occurs at the start of the path, where the Kalman filter does not need to process many points to get to the most recent estimates. Avoiding filtering grade estimates for waypoints that have been passed by storing the Kalman filter's state in memory can minimize the number of points the Kalman filter needs to process upon each iteration of this system.

\section{Conclusion}
We compared the performance of an inertial map-based and a lidar-based road grade estimation system through field experiments. The lidar-based system measures the height of the road at the points of contact for each wheel at every waypoint in the oncoming path. Odometry bias and uncertainty in each measurement are filtered through a linear correction step and a Kalman filter. From testing in a closed course on highway roads, the lidar-based approach is shown to have unbiased error within a standard deviation of $0.6^\circ$ relative to a map generated from inertial measurements. This precision enables a lidar-based measurement source to provide a preview signal of the oncoming road grade. Using this system on board an AV or an Advanced Driver Assistance System (ADAS)-equipped vehicle provides another measurement source for tasks involving grade information.


\bibliographystyle{ieeetr} 
\bibliography{bibtex/references}

\section{Contact Information}
Logan Schexnaydre \newline
lpschexn@mtu.edu

\section{Acknowledgments}
This work was supported by the ARPA-E NEXTCAR Program (Award number: DE-AR0000788). 

\section{Definitions, Acronyms, Abbreviations}

\begin{table}[h]
\centering
\begin{tabular}{L{0.1\textwidth} L{0.33\textwidth}}
\textbf{ADAS} & Advanced Driver Assistance System \\
\textbf{AV} & Autonomous Vehicles \\
\textbf{DEM} & Digital Elevation Maps \\
\textbf{GNSS} & Global Navigation Satellite System \\
\textbf{INS} & Inertial Navigation System \\
\textbf{PCA} & Principal Component Analysis
\end{tabular}
\end{table}



\end{document}